\documentclass[12pt]{article}

\usepackage{graphicx}
\usepackage{bm}
\usepackage{epstopdf}

\begin{document}

\title{A pairing hypothesis based on resonating valence bond state
for hole doped copper oxide high temperature superconductors}

\author{Ming-Lun Chen$^{1}$ \\
The email of corresponding author: mlcen@126.com \\
ORCID: 0000-0001-7578-4119 \\
\normalsize{${}^{1}$Department of Physics, Jinggangshan University, Ji'an 343009, China} \\}

\maketitle

\begin{abstract}

To explain the superconductivity of hole-doped copper-oxide
high-temperature superconductors (HDCO-HTSCs), Anderson proposed
a theory: (A) the pseudogap state is a resonating valence-bond (RVB)
state below $T^{*}$ and (B) the RVB state translates itself into
high-temperature superconducting state below $T_{c}$. In this paper
we retain Anderson theory A, add three new hypotheses and construct
an effective Hamiltonian of HDCO-HTSCs. From the effective Hamiltonian
we obtain a relationship between $T_{c}$ and hole-doping concentration
$x$ and find the trend of $T_{c}$-line is consistent with that of
underdoping region phase diagram of HDCO-HTSCs.

\end{abstract}


\maketitle

\section{Introduction}

For conventional superconductors, crystal lattice is a background of
superconducting state and its Goldstone modes (phonons) provide
glue for Cooper pairs. Although experiments \cite{1,2} shows that
Cooper pairs are still carries of hole-doped copper-oxide high-temperature
superconductors (HDCO-HTSCs), the breakthrough of McMillan limit reveals
that the glue of Cooper pairs of HDCO-HTSCs is no longer provided by
crystal lattice background. If high-temperature superconducting state
still need one background to provide glue for its Cooper pairs,
what is the kind of background? In this paper we propose this
background is a resonating valence-bond (RVB) state \cite{3}
which Goldstone modes (rubions) provide glue for Cooper pairs
of high-temperature superconducting state.

The proposal that RVB state is a background of high-temperature
superconducting state is supported by the phenomenon \cite{4} that
pseudogap coexists with superconducting gap. Anderson theory
A \cite{3}--- pseudogap state is a RVB state--- still provides a good
explanation with pseudogap phenomenon, in which pseudogap comes
from pair-breaking of singlet pair and is a high energy single particle
excitation of RVB state. So pseudogap is a signal of RVB
state same as superconducting gap is a signal of superconducting
state. The phenomenon that pseudogap coexists with superconducting
gap reveals that RVB state coexists with high-temperature superconducting
state and the former is a background of the latter below $T_{c}$.

There are two kinds of pictures about the phenomenon that pseudogap
coexists with superconducting gap. One is single-gap picture
\cite{5,6,7,8} and another is two-gap picture
\cite{9,10,11,12,13,14}. Single-gap picture shows that pseudogap is a
precursor of superconducting gap, which supports Anderson theory A
and simultaneously supports Anderson theory B \cite{15}: below $T_{c}$
the RVB state translates itself into high-temperature superconducting
state in which singlet pairs melt into Cooper pairs. However, two-gap picture
shows that pseudogap is different essentially from superconducting gap
and Cooper pairs can't come from the melting of singlet pairs, which
rejects Anderson theory B but still supports Anderson theory A. This
poses a rather puzzling situation and has been extensively discussed
\cite{16}.

Because single- and two-gap picture together support Anderson theory
A, one way out of the dilemma is to abandon Anderson theory B but
still to retain Anderson theory A and to add new hypotheses. In this
paper, we retain Anderson theory A, pseudogap state is a RVB state, and
add three new hypotheses as following: (1) RVB state is a
background of high-temperature superconducting state. (2) The RVB
background possesses Goldstone modes which are glue of Cooper pairs.
(3) Cooper pairs don't come from the melting of singlet pairs
but have other origin. Based on above three hypotheses, we construct
an effective Hamiltonian of HDCO-HTSCs, and from this Hamiltonian
we obtain a function relationship between $T_{c}$ and doping
concentration $x$ and find the trend of $T_{c}$-line is
consistent with that of underdoping region phase diagram
of HDCO-HTSCs.

It always needs to introduce some new concepts to construct
an effective Hamiltonian. In this paper we introduce three
new concepts as following: (1) ``RVB-background", i.e. RVB state,
which is a background of high-temperature superconducting state.
(2) ``Rubions", Goldstone modes of RVB-as-a-background, which are
glue of Cooper pairs. (3) ``free-$d$-electrons", electrons that
make up Cooper pairs, which come from the evolving of the
$d_{x^{2}-y^{2}}$ electrons in process of hole-doping.

This paper is organized as follows. In Sec. 2 three new hypotheses are
presented in details. In Sec. 3 based on three hypotheses we construct an
effective Hamiltonian of HDCO-HTSCs. In Sec. 4 we calculate the relationship
between $T_{c}$ and hole-doping concentration $x$ in underdoping
region, and explain the cause of high critical temperature $T_{c}^{max}$
and why $T_{c}$-line is a dome. Finally, in Sec. 5, we discuss three
questions: (1) What is the rubion? (2) What does define high temperature
superconductivity? (3) How to prove our theory false?

\section{Hypotheses}

\subsection{RVB state is merely a background}

Recently, two experiments together imply a phase-transition taking
place at $T^{*}$. One experiment \cite{17} shows that pseudogap phase is
not a crossover region but an independent phase precisely linked with
$T^{*}$. Another experiment \cite{18} shows that onset of
pseudogap phase is abrupt when temperature traverses $T^{*}$. Above
two experiments together imply that a phase-transition takes place
at $T^{*}$ and a new electronic state hides itself in the pseudogap
phase. The findings of two experiments also imply that Anderson
theory A---the pseudogap state is a RVB state---is reasonable,
in which the RVB state is just the hidden electronic state and
$T^{*}$ is just the phase-transition temperature.

There is a further proof about the propose that RVB state is a background
of superconducting state. Recently,the picture that RVB state coexists
with high-temperature superconducting state is supported strongly by that
pseudogap exists clearly in the overdoping region \cite{19}. If two states
coexist, it means that one state lies likely in bottom and becomes a background
of another state. We suppose that RVB state is located at bottom and
is a background of superconducting state below $T_{c}$. Thus,
high-temperature superconducting state has two backgrounds: one is
crystal lattice background and another is the RVB background.

The hypothesis of RVB background can explain an experiment
observation. Zheng \textsl{et al.} \cite{20}, found pseudogap
still exists in the $T \rightarrow 0$ limit when a strong magnetic
field destroys high-temperature superconducting state. According to
Anderson theory A, pseudogap is a signal of RVB state existence. If
RVB state is a background and located at bottom of superconducting
state below $T_{c}$, it can explain Zheng \textsl{et al.}'s experiment
observation.

\subsection{The RVB background possesses Goldstone modes}

According to Goldstone theorem, if continuous symmetries are broken,
system will possess a zero-mass collective exciting Goldstone mode.
The crystal lattice background breaks continuous translational and
rotational symmetries in real-space, so it possesses a zero-mass
Goldstone mode: phonon, which is glue of conventional superconductors.
If RVB state is another background of high-temperature superconducting
state, whether does it also break continuous symmetries and possess a
zero-mass Goldstone mode to provide glue for HDCO-HTSCs?

Kohsaka \textsl{et al.} \cite{1}, looked into the behavior of
electrons of an underdoped cuprate in real and in reciprocal
space simultaneously by scanning tunneling microscopy; they found
that the pseudogap excitations, locally at atomic scale, are real
space excitations that lack the delocalized characteristics. Pseudogap
excitations come from pair-breaking of singlet pairs and are the high
energy single particle excitations of RVB state. The fact that pseudogap
excitations lack the delocalized characteristics reveals that singlet
pairs localize at copper site of copper-oxide plane so that RVB state
is a valence-bond solid \cite{15} of real space and rules out a possibility
that the pseudogap state is a charge density wave state \cite{21}.

If RVB state is a valence-bond solid of real space, same as crystal
lattice, it will break continuous translational and rotational symmetries
in real space. Breaking of continuous symmetries leads to the emergence of a
zero-mass Goldstone mode according to Goldstone Theorem. We name the
zero-mass collective exciting Goldstone mode as ``rubion". And since
the rubion is massless, it can induce long-range interaction and serve
as the glue of HDCO-HTSCs.

The rubion hypothesis can explain an experiment observation.
Doiron-Leyraud \textsl{et al.} \cite{22}, found a nonzero thermal
conductivity for underdoped $YBa_{2}Cu_{3}O_{y}$ in the $T
\rightarrow 0$ limit. It was attributed by Doiron-Leyraud \textsl{et al.}
to a contribution of a new boson mode. If rubion is just the new boson
mode, it can explain Doiron-Leyraud \textsl{et al.}'s experiment
observation.

Through exchanging a rubion, two nonlocalized electrons are bonded
into one Cooper pair. However, even if this bonding way is true,
where do the nonlocalized electrons come from?

\subsection{The nonlocalized electrons which make up Cooper pairs
come from the evolving of the $d_{x^{2}-y^{2}}$ electrons}

To explain the origin of Cooper pairs, Anderson theory B \cite{15}
proposed that RVB state translates itself into  high-temperature
superconducting state when cuprates are doped so sufficiently that
singlet pairs melt into Cooper pairs. However, the proposal that
Cooper pairs come from singlet pairs is rejected by two-gap picture
and further rejected by Zheng \textsl{et al.}'s experiment \cite{20}:
when a strong magnetic field destroys high-temperature superconducting state,
pseudogaps still exist in the $T\rightarrow 0$ limit. The existence
of pseudogaps shows singlet pairs still exist when Cooper pairs is
killed by a strong magnetic field. Zheng \textsl{et al.}'s experiment is not in
conflict with Anderson theory A (RVB theory) but shows that it is
unlikely that Cooper pairs come from the melting of singlet pairs
and Anderson theory B must be abandoned. We need to find a new
origin about Cooper pairs.

Copper-oxide plane is a conducting layer and $d_{x^{2}-y^{2}}$
electrons (for short, $d$ electrons) of copper site are responsible
for the superconductivity. According to Anderson theory A, RVB state
seems to be participated by all $d$ electrons. If retains Anderson
theory A, on the copper-oxide plane what else electrons make up
Cooper pairs?

In this paper, Anderson theory A is improved by us as following:
pseudogap state is a RVB state but it is not all $d$ electrons to
participate in RVB state due to hole dopant. In hole-doping process,
holes don't enter into copper sites but into oxygen sites. These
holes into oxygen sites makes $d$ electrons of copper sites
evolve into two categories: one category participating in RVB state
at the $T^{*}$ and another evolving into Cooper pairs below the
$T_{c}$, i.e. it is not all $d$ electrons to participate in RVB
state.

Why in the hole-doping process $d$ electrons don't totally
participate in RVB state but evolve into two categories? (1) RVB
state is a linear superposition of singlet pairs and the onset of
one singlet pair needs a vital condition that is the superexchange
(or Kramers-Anderson superexchange \cite{23}). (2) Oxygen ions are
nonmagnetic in copper-oxide plane and nonmagnetic oxygen ions are
superexchange media of singlet pair. A hole enters into an oxygen
site to make a nonmagnetic oxygen ion become a magnetic oxygen ion;
it means that a superexchange medium is destroyed and a singlet pair
is broken up and two $d$ electrons are set free. (3) The more holes
are doped into oxygen site, the more superexchange media are
destroyed, and the more $d$ electrons are free outside from RVB
state, which are named as ``free-$d$-electrons" by us. These
free-$d$-electrons are just the original electrons of
Cooper pairs.

The hypothesis of free-$d$-electrons can explain following two
experiments: (1) Sun \textsl{et al.} \cite{24}, found that
delocalized fermions exist in underdoped $YBa_{2}Cu_{3}O_{y}$. If
these delocalized fermions are just the free-$d$-electrons,
it can explain Sun \textsl{et al.}'s experiment observation.
(2) Doiron-Leyraud \textsl{et al.} \cite{25}, measured quantum
oscillations in an underdoped cuprate. They found that the quantum
oscillation signals occurring in a Hall coefficient which has a
negative sign. The Hall coefficient is expected to be positive
according to conventional ideas (holes are thought as carries
of HDCO-HTSCs). However, the Hall coefficient is negative in
Doiron-Leyraud \textsl{et al.}'s experiment observations.
It means that the carries of HDCO-HTSCs are free-$d$-electrons
instead of the holes.

\section{Hamiltonian}

We can construct an effective Hamiltonian of HDCO-HTSCs based on
above three hypotheses---rubion, RVB background and
free-$d$-electrons. When one free-$d$-electron goes through the RVB
background, it causes a deformation of RVB background, i.e. one
free-$d$-electron emits a rubion. When another free-$d$-electron
walks into the deformed region, it feels an attraction, i.e. another
free-$d$-electron absorbs a rubion. Above process can be clarified
by a Feynman Diagram as Fig. 1. This process is not a direct Coulomb
interaction but a rubion-induced electron-electron attractive
interaction which can be presented by following formula
\begin{figure}[tbp]
\includegraphics[width=0.5\textwidth]{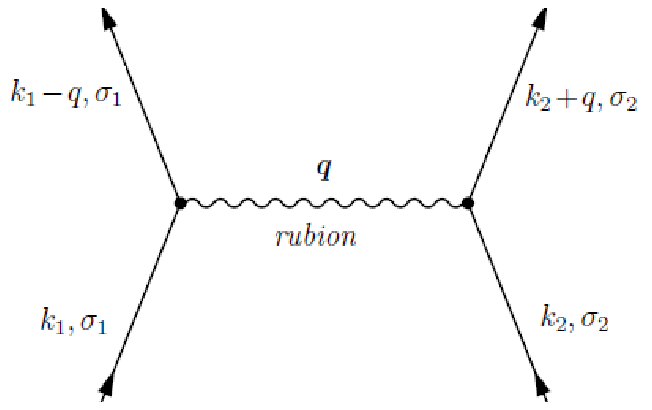}
\caption{Rubion-induced electron-electron interaction. The
$k_{1},\sigma_{1}$ is the first free-$d$-electron which becomes
$k_{1}-q,\sigma_{1}$ after emits a rubion q, and the
$k_{2},\sigma_{2}$ is the second free-$d$-electron which becomes
$k_{2}+q,\sigma_{2}$ when absorbs a rubion q. Through media of a
rubion q, two free-$d$-electron build up an attractive interaction.}
\end{figure}
\begin{equation}
H_{rubion}=\frac{1}{2}\sum\limits_{q,k_{1},k_{2},\sigma_{1},\sigma_
{2}}R_{k_{1},k_{2},q}C_{k_{1}-q,\sigma_{1}}^{\dagger
}C_{k_{2}+q,\sigma_{2}}^{\dagger}C_{k_{2},\sigma_{2}}C_{k_{1},\sigma_{1}},
\end{equation}
where $R_{k_{1},k_{2},q}$ is a rubion-induced attractive potential
and $\textbf{\textit{q}}$ represents a rubion.

In addition, there is a direct Coulomb interaction between two
free-$d$-electrons. According to Quantum Electrodynamics, this
direct Coulomb interaction should be presented by exchanging photon
between two free-$d$-electrons. However, because of Coulomb screen
of RVB state (most $d_{x^{2}-y^{2}}$ electrons participate in RVB
state), the direct Coulomb interaction between two
free-$d$-electrons can be presented by an effective interaction,
\begin{equation}
H_{coul}=\frac{1}{2}\sum\limits_{q,k_{1},k_{2},\sigma_{1},\sigma_
{2}}U_{k_{1},k_{2},q}C_{k_{1}-q,\sigma_{1}}^{\dagger
}C_{k_{2}+q,\sigma_{2}}^{\dagger}C_{k_{2},\sigma_{2}}C_{k_{1},\sigma_{1}},
\end{equation}
where $U_{k_{1},k_{2},q}$ is a RVB Coulomb screening repulsive
potential and $\textbf{\textit{q}}$ still represents a rubion.

Combining above two interaction, the effective Hamiltonian of
HDCO-HTSCs is written as
\begin{equation}
H=H_{rubion}+H_{coul}.
\end{equation}
If rubion-induced attractive potential $R_{k_{1},k_{2},q}$ is
greater than the RVB Coulomb screening repulsive potential
$U_{k_{1},k_{2},q}$, then $R_{k_{1},k_{2},q}+U_{k_{1},k_{2},q}$ will
be a net attractive potential and two free-$d$-electrons can be
sticked into one Cooper pair.

According to formula (3), if only considering those interactions
which scatter a pair of free-$d$-electrons of opposite momentum and
spin $(k\uparrow, k\downarrow)$ to another pair state $(k'\uparrow,
k'\downarrow)$, the interactions take the simplified form:
\begin{equation}
H_{rubion}=\frac{1}{2}\sum\limits_{kk'}R(k-k')C_{k'}^{\dagger
}C_{-k'}^{\dagger}C_{-k}C_{k},
\end{equation}
\begin{equation}
H_{coul}=\frac{1}{2}\sum\limits_{kk'}U(k-k')C_{k'}^{\dagger
}C_{-k'}^{\dagger}C_{-k}C_{k}.
\end{equation}
On the Fermi arc/pocket \cite{26} of reciprocal space, we can
introduce an averaged strength for the net electron-electron
interaction,
\begin{equation}
-V=\langle-R(k-k')+U(k-k')\rangle_{Av},
\end{equation}
where $V$ is positive and has a $d$-wave symmetry \cite{27} and the
average is to be carried out over all free-$d$-electrons on the
Fermi arc/pocket of reciprocal space. With the help of $V$, formula
(3), can be presented as
\begin{equation}
H=-\sum\limits_{kk'\in arc/pocket}VC_{k'}^{\dagger
}C_{-k'}^{\dagger}C_{-k}C_{k}.
\end{equation}
Formula (7) is the effective Hamiltonian of HDCO-HTSCs.

\section{Results}

From formula (7), we can find out that the effective Hamiltonian of
HDCO-HTSCs is still a BCS-like Hamiltonian, in which the only
difference is that $V$ represents the electron-rubion interaction
instead of electron-phonon interaction.

With the aid of $V$, we can define the energy gap,
\begin{equation}
\Delta=V\sum\limits_{k}\langle C_{-k}C_{k} \rangle,
\end{equation}
\begin{equation}
\Delta^{*}=V\sum\limits_{k}\langle C_{-k}^{\dagger}C_{k}^{\dagger}
\rangle,
\end{equation}
where both $\langle C_{-k}C_{k} \rangle$ and $\langle
C_{-k}^{\dagger}C_{k}^{\dagger} \rangle$ are pair operator and the
average is to be carried out over superconducting ground state which
is a distribution of Cooper pairs on the Fermi arc/pocket of
reciprocal space.

Wang \textsl{et al.}'s experiment \cite{28} shows that the weak-coupling
$d$-wave BCS universal relation,
\begin{equation}
\Delta_{0}=2.14k_{B}T_{c},
\end{equation}
still exists in HDCO-HTSCs and even in overdoping region. It reveals
that the rubion-induced electron-electron interaction is still a
weak-coupling interaction. It means the weak-coupling approximation,
$N_{0}V\ll1$, can be used for following deducing process.

Under the weak-coupling approximation, starting from formula (8, 9)
and repeating same deducing procedure as weak-coupling BCS theory
\cite{29}, we obtain an expression of high-temperature
superconducting energy gap in $T=0K$ as following
\begin{equation}
\Delta_{0}=2\hbar\omega_{D}exp[-1/N_{0}V],
\end{equation}
where $\omega_{D}$ is Debye frequency of rubions and $N_{0}$ is the
energy state density of free-$d$-electrons on the Fermi arc/pocket.
For HDCO-HTSCs, $\Delta_{0}$ is $10^{-2}eV$ magnitude according to
experimental data \cite{4}. If we set $N_{0}V$ as 0.1 based on the
weak-coupling limit $N_{0}V\ll1$, then the Debye temperature
$\Theta_{D}$($\Theta_{D}=\hbar\omega_{D}/k_{B}$) of rubions is equal
to $10^{3}K$ magnitude approximately.

\begin{figure}[tbp]
\includegraphics[width=0.5\textwidth]{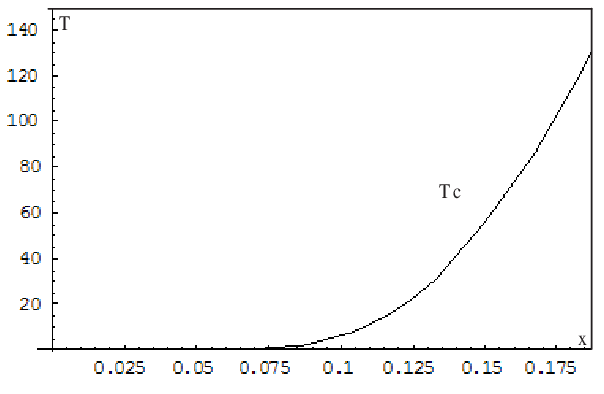}
\caption{$T_{c}$ as a function of doping concentration $x$ according
to function relationship $T_{c}=0.93\Theta_{D}exp(-\frac{1}{V\alpha
x})$, which is derived from the effective Hamiltonian of HDCO-HTSCs.
$\Theta_{D}$ is set as $5\times10^{3}K$, $\alpha$ as 15, and $V$ as
0.1. The trend of $T_{c}$-line is consistent with that of underdoping
region phase diagram of HDCO-HTSCs.}
\end{figure}

Comparing formula (10) with (11), we obtain the $T_{c}$ expression
of HDCO-HTSCs as following
\begin{equation}
T_{c}=0.93\Theta_{D}exp(-1/N_{0}V),
\end{equation}
Further, according to the hypothesis which free-$d$-electrons are
induced by hole-doping process, the $N_{0}$ should be proportional
to hole-doping concentration $x$, i.e. $N_{0}=\alpha x$ where
$\alpha$ is scale factor. In formula (12), substituting $N_{0}$ with
$\alpha x$, we can obtain a function relationship between $T_{c}$ and
$x$ as following
\begin{equation}
T_{c}=0.93\Theta_{D}exp(-\frac{1}{V\alpha x}),
\end{equation}
where $V\alpha x\ll 1$ is required by the weak-coupling limit. Based
on formula (13), the function plots in underdoping region is
presented as Fig. 2 in which the trend of $T_{c}$-line is consistent
with that of underdoping region phase diagram of HDCO-HTSCs.

\section{Understanding}

\subsection{Cause of a dome-shaped $T_{c}$-line}

In phase diagram of HDCO-HTSCs, $T^{*}$-line reflects an abilities
of RVB state withstanding thermal-fluctuation. The ability is
gradually weaken by quantum fluctuation which is imported by
hole-dopant. So, $T^{*}$-line stretches inevitably towards lower
right of phase diagram. On the other hand, $T_{c}$-line reflects
condensation energy of high-temperature superconducting state.
Based on the free-$d$-electron hypothesis, the more holes are doped,
the more free-$d$-electrons are free out. It means the more Cooper
pairs are made up and the stronger phase stiffness is owned by
high-temperature superconducting state with increasing hole-doping
concentration. So $T_{c}$-line stretches inevitably towards the
upper right of phase diagram until meets $T^{*}$-line (see Fig. 3).

\begin{figure}[tbp]
\includegraphics[width=0.5\textwidth]{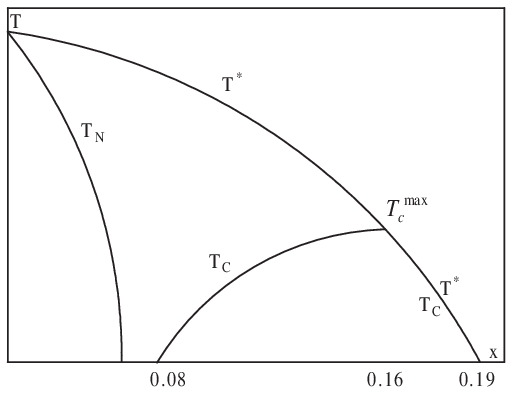}
\caption{Rubion hypothesis predicts that the optimal dopant point is
located at the meeting point of $T_{c}$-line and $T^{*}$-line. After
the optimal dopant point, $T_{c}$-line blends into $T^{*}$-line and
together extend to $x=0.19$ where an experiment \cite{30} shows that
pseudogap only exists in $x\leq0.19$.}
\end{figure}

Based on the rubion hypothesis, RVB state is a precondition of
high-temperature superconducting state, so that, after the meeting
point, the $T_{c}$-line blends inevitably into the $T^{*}$-line and
the $T^{*}$-line becomes a precondition of $T_{c}$-line. Once RVB
state is ruined by the thermal fluctuation, high-temperature
superconducting state collapses instantly because that glue of the
latter is provided by the collective modes of the former. So in
overdoping region the $T_{c}$-line blends inevitably into the
$T^{*}$-line and whole $T_{c}$-line cannot be anything but a dome
(see Fig. 3). In Fig. 3 the optimal dopant point is located at the
meeting point of $T^{*}$-line and $T^{*}$-line, and $T_{c}^{max}$
is just equal to $T^{*}$ at optimal dopant point.

\subsection{Cause of a high critical temperature $T_{c}^{max}$}

Because Wang \textsl{et al.}'s experiment \cite{28} reveals that
the carries of high-temperature superconducting state are still
weak-coupling Cooper pairs, even if rubion is the glue of Cooper
pairs, the rubion-induced electron-electron attracting interaction
is not more prominent than the phonon-induced electron-electron
attracting interaction. Why does rubion-induce high-temperature
superconducting state have a higher critical temperature $T_{c}^{max}$
at optimal dopant point than that of conventional superconductors?

Firstly we analysis how thermal fluctuation breaks up Cooper pairs
of conventional superconductors. For conventional superconductors,
the glue of Cooper pairs comes from phonon-induced electron-electron
attracting interaction which is achieved through an electron-induced
local distortion of crystal lattice background. When temperature
arising, thermal fluctuation increasingly smoothes the local distortion
of crystal lattice background and then Cooper pairs lose their glue.
So it is a low order effect that thermal fluctuation breaks up
phonon-induced Cooper pairs and the $T_{c}$ of conventional
superconducting state can not break through McMillan limit.

For HDCO-HTSCs, superconducting state has two backgrounds: one is
crystal lattice background and another is the RVB background.
Based on rubion hypothesis, the glue of Cooper pairs is provided
by RVB background which is achieved through a free-$d$-electron-induced
local distortion of RVB background. If thermal fluctuation wants to break up
rubion-induced Cooper pairs through smoothing local distortion
of RVB background, it must take two steps: firstly thermal fluctuation
perturbs crystal lattice background, and secondly, through scattering
interaction of crystal lattice background, perturbs the RVB background;
the perturbation of RVB background will smooth free-$d$-electron-induced
local distortion of RVB background and then Cooper pairs lose their glue.
Because this process needs two steps, it is a high order effect of thermal
fluctuation to break up rubion-induced Cooper pairs.

The high order effect is the inherent cause of a high critical temperature
$T_{c}^{max}$ so that the high-temperature superconducting state does not
need a more sticky to bind its cooper pairs. At optimal dopant point, due to
high order effect, it is not an easy task that thermal fluctuation causes
rubion-induced Cooper pairs losing their glue through the way of smoothing
local distortion of RVB background. The energy saving way is that the thermal
fluctuation directly destroys RVB background and then rubion-induced Cooper
pairs lose their glue source. So, the $T_{c}^{max}$ depends on the value
of $T^{*}$ at optimal dopant point and a high $T^{*}$ at optimal dopant point
leads to a high critical temperature $T_{c}^{max}$.

Although the inherent cause of a high $T_{c}^{max}$ lies in the high order
effect, the value of $T^{*}$ at optimal dopant point limits the height of
$T_{c}^{max}$. So a different series of hole doped copper oxide has a
different $T^{*}$-line so that it owns a different $T_{c}^{max}$.

On whole dome-shaped $T_{c}$-line of Fig. 3, due to high order effect, direct
pair-breaking of Cooper pairs is not an energy saving way for thermal fluctuation
destroying high-temperature superconducting state. Before optimal dopant point,
destroying phase condensation of Cooper pairs is an energy saving way so that
Cooper pairs can exist above $T_{c}$-line, which is the cause that Nernst effect
\cite{31} can exist between $T_{c}$-line and $T^{*}$-line in underdoping region.
At and After optimal dopant point, destroying RVB state is another energy saving
way, in which Cooper pairs lose their glue source. The existence of two destroying
ways is the inherent cause of $T_{c}$-line being a dome.

\subsection{Cause of the hydrostatic pressure promoting $T_{c}^{max}$}

The hydrostatic pressure brings about two factors for promotion of $T_{c}^{max}$.
One is holes increasing caused by oxygen ordering effects \cite{32}. The increase of
holes means that the superexchanging mediums decrease and free-$d$-electrons increase.
The increase of free-$d$-electrons means that Cooper pairs increase and phase stiffness
strengthens and $T_{c}^{max}$ ascends. Another is the superexchanging interaction $J$
increasing \cite{33}. At the optimal dopant point, the increase of $J$ means the RVB state
becomes more and more stable so that the ascended $T_{c}^{max}$ can be beared by the
RVB background. Similarly the multi-layer copper oxide planes under hydrostatic pressure
further reinforces the stability of RVB background due to the coupling between layers
so that a higher $T^{*}$ at the optimal dopant point leads to a higher $T_{c}^{max}$.

\subsection{Cause of the deviation between $T_{c}$-line and condensation
energy in overdoping region}

Loram \textsl{et al.}'s experiment \cite{34} shows condensation energy
is in proportion to hole-dopant concentration even though in overdoping
region. However, in phase diagram of overdoping region, $T_{c}$-line
declines with hole-dopant concentration $x$ so that $T_{c}$-line deviates
from the trend of condensation energy.

Rubion hypothesis can explain the deviation between $T_{c}$-line and
condensation energy in overdoping region. Based on the free-$d$-electron
hypothesis, the more holes are doped, the more free-$d$-electrons are free out.
It means the more Cooper pairs are made up and the stronger phase stiffness is
owned by high-temperature superconducting state with increasing hole-doping
concentration. This is the cause that condensation energy is in proportion to
hole-dopant concentration in overdoping region. Again based on the rubion
hypothesis, the $T^{*}$-line becomes a precondition of $T_{c}$-line so that
the $T_{c}$-line must blend into the $T^{*}$-line in overdoping region. So,
in overdoping region $T_{c}$-line deviates inevitably from the trend of
condensation energy.

\section{Discussions}

\subsection{What is the rubion?}

In this paper, we theoretically examined the phase diagram of cuprate
high-temperature superconductors based on three hypotheses. The most
crucial hypothesis is the presence of a collective mode named as rubion
by us. The mode is assumed to be generated from the RVB state as a
Goldstone mode. In order to have the Goldstone mode, one has to have a
long-range order characterized by a corresponding order parameter.
Kohsaka \textsl{et al.}'s experiment \cite{1} reveals RVB state is
a valence-bond solid of real space so that the long-range order is a
positional order and the corresponding order parameter is $\rho_{G}$ \cite{35}.

Then, what is the rubion? Firstly, rubion can not be a plasmon because
plasmon is a collective excited mode of wide-band metal while
HDCO-HTSCs are narrow-band bad-metals. Second, rubion can not be a
spin wave because spin wave is a Goldstone mode of ferromagnetism or
antiferromagnetism while RVB state recover spin rotation symmetry
when hole-dopant process kills antiferromagnetism of parent
compound. So rubion is a new low-energy collective excited mode.

If rubion is a new collective mode, what is the properties of
rubion? Firstly, as a Goldstone mode, rubion ought to be a spinless
boson of zero mass according to Goldstone theorem. Second, as a
quantum of collective mode, rubion ought to possess an energy
$\hbar\omega$, where $\omega$ is frequency of new collective mode.
Rubion also ought to contribute a thermal conductivity to HDCO-HTSCs
and its Debye temperature is equal  approximately to $10^{3}K$ magnitude
according to the calculation of section Applications.

Doiron-Leyraud \textsl{et al.}'s experiment \cite{22} finds out
a nonzero thermal conductivity for underdoped $YBa_{2}Cu_{3}O_{y}$
in the $T \rightarrow 0$ limit, which was attributed to a contribution
of a new boson mode. If rubion is just the new boson
mode, this experiment can determine the dispersion relation
$\omega=\omega(q)$ of rubion.

\subsection{What does define high temperature superconductivity?}

For conventional superconducting state, the perturbation of thermal
fluctuation of crystal lattice background is so direct that the $T_{c}$
can not breaks through McMillan limit. However, for high-temperature
superconducting state, there is an interlayer between superconducting
state and crystal lattice background, and the interlayer is just RVB
state which leads to the pair-breaking of Cooper pairs is a high order
effect of thermal fluctuation. So, about the question: ``What does define
high temperature superconductivity? \cite{36}", our answer is as following:
due to existance of RVB interlayer, that thermal fluctuation breaking up
rubion-induced Cooper pairs is a high order effect, defines high temperature
superconductivity.

\subsection{How to prove rubion hypothesis false?}

According to the proposal of Karl Popper, if a theory belongs to
science, it must own a falsifiability.

Rubion hypothesis comes to a conclusion that the $T^{*}$-line is
a precondition of $T_{c}$-line, which rules out Fig. 4 as an
universal phase diagram \cite{37}. In Fig. 4, the character that
Cooper pairs still exist above $T^{*}$-line in overdoping region
reminds that experimental scientists can design an experiment to
check the falsifiability of our theory. This is an experiment based
on Nernst effect because the Nernst effect can reflect the existing
range of Cooper pair \cite{38}. Because Fig. 4 rejects rubion
hypothesis, if experimental scientists design the experiment to find
out Nernst effect existing above $T^{*}$-line in overdoping region,
it will prove our theory false.

Recently, Chang \textsl{et al.}'s experiment \cite{39} shows that
there is a controversy about the range of Nernst effect but the
controversial range is below the $T^{*}$-line in underdoping region,
in which the boundary of forecast of rubion hypothesis isn't overflowed.
Howerver, for overdoping region, there are too few experimental data and
too much controversy such as recently Vishik \textsl{et al.}'s experiment
\cite{19} gives a different pseudogap evolution way. Similarly, rubion
hypothesis predicts that the $T_{c}$-line blends into the $T^{*}$-line in
overdoping region (see Fig. 3). Rubion hypothesis will further stimulate
experimental scientists to pay close attention to overdoping region.

\begin{figure}[tbp]
\includegraphics[width=0.5\textwidth]{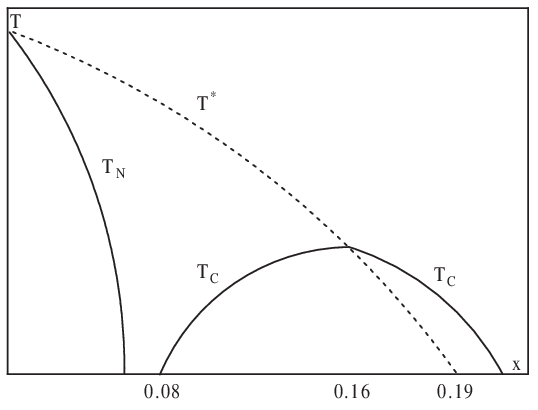}
\caption{Rubion hypothesis rules out a possibility that Fig. 4 is as
an universal phase diagram. Fig. 4 comes from Fischer \cite{37}
\textsl{et al.}}
\end{figure}

\subsection{What new physics does rubion hypothesis reveal?}

Time went on with its work twenty-seven years since HTSCs were discovered.
Scientists waited in hope for a new physics from HTSCs as the same experience
that Laughlin brought us from the fractional hall effect. However, rubion
hypothesis shows that there is not a new physics existing in HTSCs and
if HTSCs still own Cooper pairs as its carries \cite{1,2}, its physics
continues to be BCS, although most scientists are unwilling to accept a
trival conclusion.

If we stubbornly intend to find out a few new physics from HTSCs, the few new
physics lies first in that HTSCs own a peculiar electron state (RVB), and
then the peculiar electron state has an elementary excitation (rubion). The
peculiar electron state can explain the origin of $T^{*}$-line. The elementary
excitation can explain the origin of the dome-shaped $T_{c}$-line, and predict
that the $T_{c}$-line meets the $T^{*}$-line and the meeting point is just
the optimal dopant point (see Fig. 3).

ACKNOWLEDGMENTS. We thank Ruihua He of Boston college for fruitful
discussions. This work was financially supported by the NSFC (Grant
No.11164010).

\end{document}